\documentclass[floatfix,aps,twocolumn,floatfix,10pt,superscriptaddress,prl]{revtex4-1}

\usepackage{bm}
\usepackage{graphicx}
\usepackage{subfigure}
\usepackage{epstopdf}
\usepackage{siunitx}
\usepackage{braket}
\usepackage{tabularx}
\usepackage{blindtext}
\usepackage{amsmath}

\usepackage{color}

\def\bem#1{\begin{mathletters}\label{#1}}
\def\eml{\end{mathletters}}

\def\ket#1{{|#1\rangle}}

\def\4#1{{\boldsymbol{#1}}}
\def\8#1{{\widetilde{#1}}}

\begin{document}

\title {Enhanced spin state readout of Nitrogen-Vacancy centers in a diamond using IR fluorescence}

\author{I. Meirzada}
\affiliation{The Racah Institute of Physics, The Hebrew University of Jerusalem, Jerusalem 91904, Israel}

\author{S. A. Wolf}
\affiliation{The Racah Institute of Physics, The Hebrew University of Jerusalem, Jerusalem 91904, Israel}
\affiliation{The Center for Nanoscience and Nanotechnology, The Hebrew University of Jerusalem, Jerusalem 91904, Israel}

\author{A. Naiman}
\affiliation{The Racah Institute of Physics, The Hebrew University of Jerusalem, Jerusalem 91904, Israel}
\affiliation{The Center for Nanoscience and Nanotechnology, The Hebrew University of Jerusalem, Jerusalem 91904, Israel}

\author{U. Levy}
\affiliation{Dept. of Applied Physics, Rachel and Selim School of Engineering, Hebrew University, Jerusalem 91904, Israel}
\affiliation{The Center for Nanoscience and Nanotechnology, The Hebrew University of Jerusalem, Jerusalem 91904, Israel}

\author{N. Bar-Gill}
\email{bargill@phys.huji.ac.il}
\thanks{Corresponding author.}
\affiliation{The Racah Institute of Physics, The Hebrew University of Jerusalem, Jerusalem 91904, Israel}
\affiliation{Dept. of Applied Physics, Rachel and Selim School of Engineering, Hebrew University, Jerusalem 91904, Israel}
\affiliation{The Center for Nanoscience and Nanotechnology, The Hebrew University of Jerusalem, Jerusalem 91904, Israel}

\begin{abstract}

Nitrogen-Vacancy (NV) centers in diamond have been used in recent years for a wide range of applications, from nano-scale NMR to quantum computation. These applications depend strongly on the efficient readout of the NV center's spin state, which is currently limited. Here we suggest a method of reading the NV center's spin state, using the weak optical transition in the singlet manifold. We numerically calculate the number of photons collected from each spin state using this technique, and show that an order of magnitude enhancement in spin readout signal-to-noise ratio is expected, making single-shot spin readout within reach. Thus, this method could lead to an order of magnitude enhancement in sensitivity for ubiquitous NV based sensing applications, and remove a major obstacle from using NVs for quantum information processing. 

\end{abstract}

\maketitle

Effective quantum state readout is a crucial component of almost every quantum computation or sensing device, and extensive research in a variety of fields is directed at improving quantum state measurements and increasing readout fidelity \cite{reed_high-fidelity_2010,myerson_high-fidelity_2008,steiner_universal_2010,morello_single-shot_2010}. The Nitrogen-Vacancy (NV) color center in diamond is a promising system for various quantum based applications, such as quantum computation \cite{fuchs_quantum_2011} and sensitive measurements \cite{clevenson_broadband_2015,acosta_broadband_2010,dolde_electric-field_2011,taylor_high-sensitivity_2008,loretz_nanoscale_2014,trusheim_wide-field_2016}, due to its unique optical and spin properties. Nevertheless, a fast and high fidelity spin state readout for the NV center is currently missing, and although extensive efforts have been invested in this context \cite{wolf_purcell-enhanced_2015,steiner_universal_2010,shields_efficient_2015,robledo_high-fidelity_2011,hopper_near-infrared-assisted_2016}, many repetitions of each measurement, cold temperatures or long measurement times are still needed for each experiment. 

This work presents a novel approach for reading the NV's spin state, based on fluorescence measurements of the singlet infrared (IR) transition. We first recalculate the standard red fluorescence based spin state readout with recently published ionization and recombination rates of NV$^-$ and NV$^0$ \cite{meirzada_negative_2017}. Next, we detail our proposed method of reading the NV center's spin state, using the weak fluorescence emitted in the singlet manifold, and calculate the expected signal-to-noise ratio (SNR) by numerically solving the relevant rate equations, for both surface and bulk NVs. From these results, we find a regime of excitation parameters for which a significant increase in the NV's spin state readout SNR is expected. Finally, we suggest using a photonic crystal cavity to increase the radiative coupling of the singlet transition, and present the quality and Purcell factors, as well as the SNR expected using these structures coupled to a nanodiamond, a diamond membrane and bulk diamond.



    
    %

The negatively charged NV center consists of 2 adjacent lattice sites occupied by a nitrogen atom and a vacancy inside a diamond crystal. The electronic ground state of the NV center is a spin triplet with a 2.87 GHz zero-field splitting between spin projections $m_s = 0$ and $m_s = \pm 1$. The electronic excited states contain a spin triplet with a strong radiative coupling and a spin singlet with a much weaker radiative coupling. 

\begin{figure}[tbh]
{\includegraphics[width=1 \linewidth]{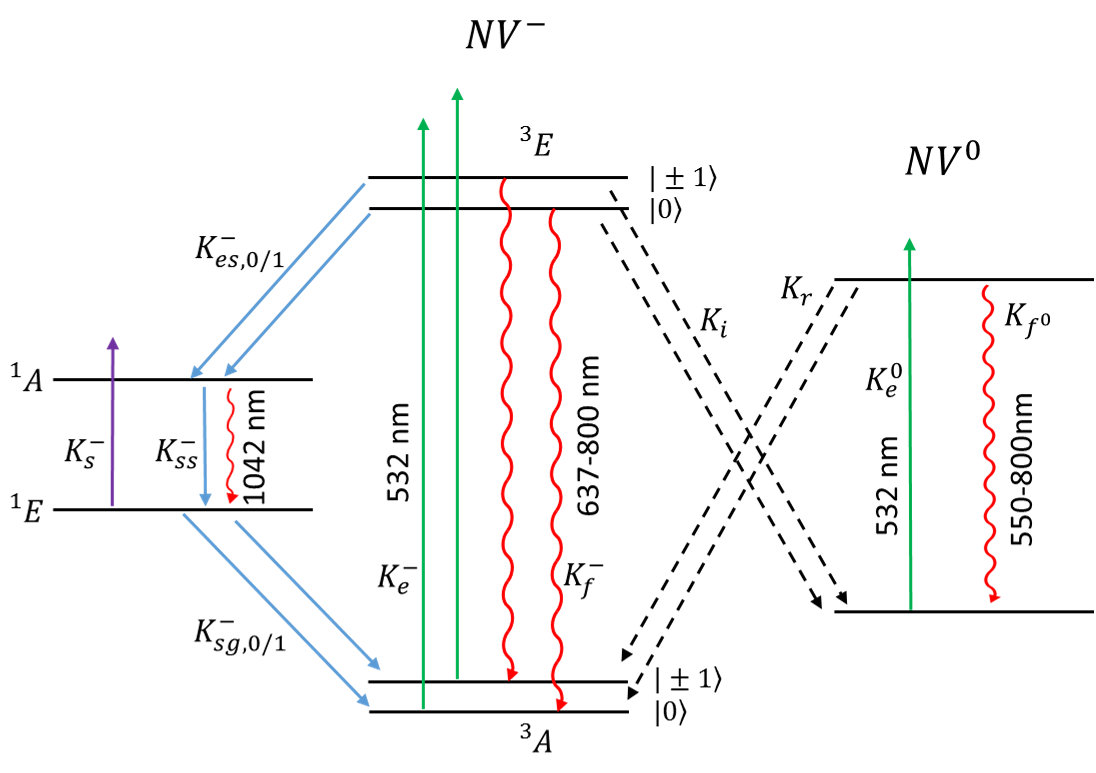}}
\caption{Energy level diagram and relevant transitions for the neutral and negatively charged NV center. Green excitation is depicted with green arrows, red fluorescence is depicted with downward red arrows, IR excitation and fluorescence are depicted with orange arrows, non-radiative decay is depicted with blue arrows, ionization and recombination transitions depicted with dashed purple arrows.} 
\label{fig:energylevel}
\end{figure}

Figure \ref{fig:energylevel} depicts a simplified energy level diagram of NV$^-$ and NV$^0$, together with their main transitions. In the standard red fluorescence spin readout scheme, an NV in the triplet ground state ($^3A$) is excited to the triplet excited state ($^3E$) using green light, and the red fluorescence during the decay back to the ground state is collected. The number of photons collected from each of the spin states dictates the SNR, which is defined under the shot noise limit assumption as:
\begin{equation} 
SNR = \frac{|N_0-N_1|}{\sqrt[]{N_0+N_1}}
\label{SNR}
\end{equation}
where $N_i$ denotes the number of photons collected when the NV is initialized to its $m_s = \ket{i}$ state, where $i$ can be 0 or 1. 

\begin{figure}[tbh]
\subfigure[]{
\includegraphics[trim = 1mm 1mm 0mm 3mm, clip, width=0.46 \linewidth]{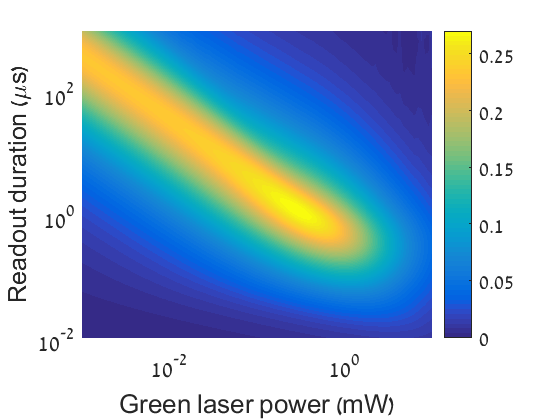}}
\subfigure[]{
\includegraphics[trim = 1mm 1mm 0mm 3mm, clip, width=0.46 \linewidth]{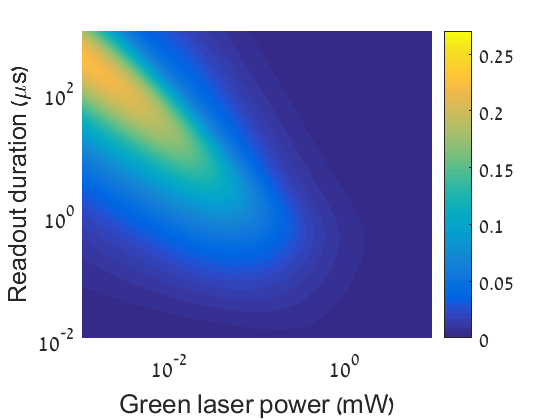}}
\subfigure[]{
\includegraphics[trim = 1mm 1mm 0mm 3mm, clip, width=0.46 \linewidth]{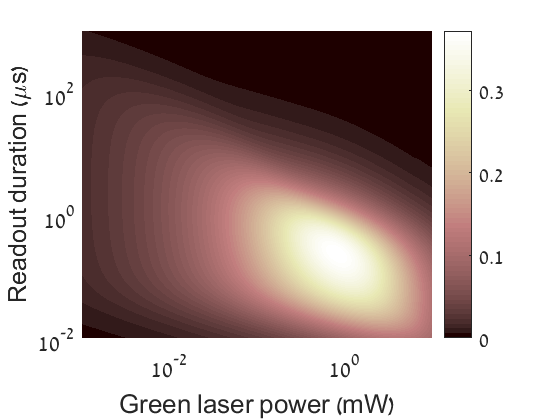}}
\subfigure[]{
\includegraphics[trim = 1mm 1mm 0mm 3mm, clip, width=0.46 \linewidth]{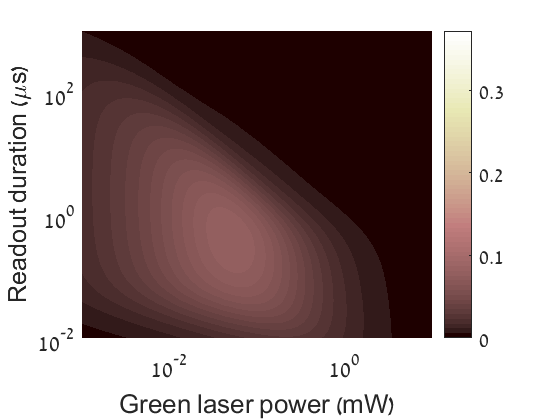}}
\protect\caption{SNR as a function of green excitation power and pulse duration for bulk (a,c) and surface (b,d) NVs, without (a,b) and with (c,d) time normalization. A big difference in the SNR between bulk and surface NVs is noticed due to the difference in ionization cross sections.}
\label{fig:greenSNR}
\end{figure}

\begin{figure}[tbh]
\subfigure[]{
\includegraphics[width=0.85 \linewidth]{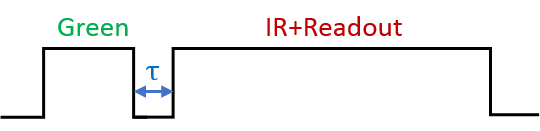}}
\subfigure[]{
\includegraphics[trim = 1mm 1mm 0mm 3mm, clip, width=0.46 \linewidth]{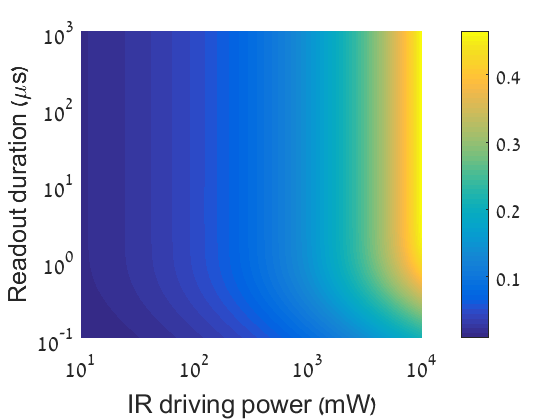}}
\subfigure[]{
\includegraphics[trim = 1mm 1mm 0mm 3mm, clip, width=0.46 \linewidth]{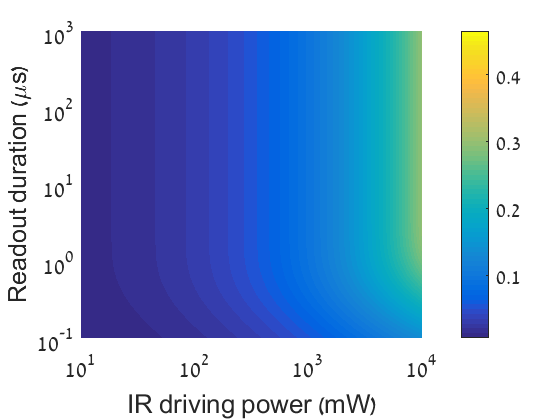}}
\subfigure[]{
\includegraphics[trim = 1mm 1mm 0mm 3mm, clip, width=0.46 \linewidth]{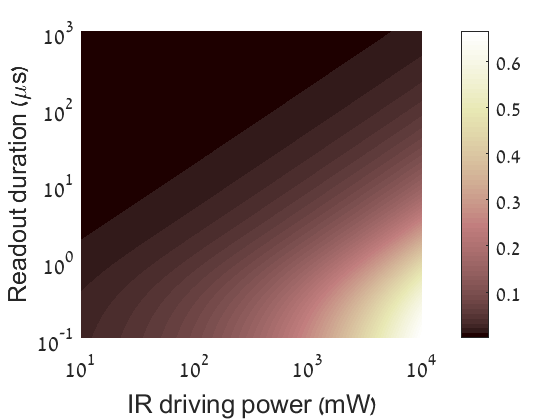}}
\subfigure[]{
\includegraphics[trim = 1mm 1mm 0mm 3mm, clip, width=0.46 \linewidth]{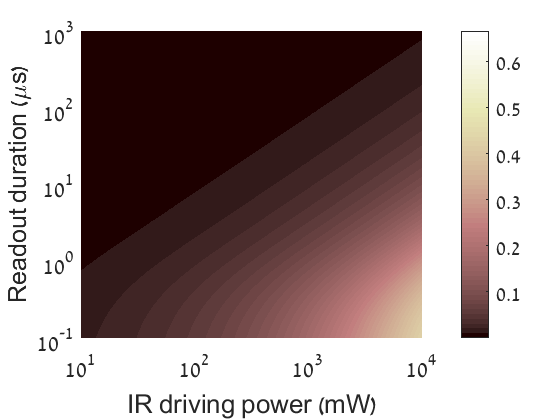}}
\protect\caption{IR fluorescence spin state readout. (a) pulse sequence for the IR fluorescence spin readout. (b-e) SNR as a function of IR excitation power and pulse duration for bulk NV (b,d) and surface (c,e) NVs, with (d,e) and without (b,c) normalization. Higher absolute and normalized SNR can be theoretically achieved compared to red fluorescence readout, for both bulk and surface NVs, using strong enough excitation.}
\label{fig:IRSNR}
\end{figure}

We first calculate the spin readout SNR using green excitation and red fluorescence detection, as a function of readout duration and excitation power for a confocal system, for both surface and bulk NVs, assuming perfect collection and detection efficiencies. In addition, fluorescence from NV$^0$ is ignored, although it overlaps to some extent with the NV$^-$ fluorescence. The SNR is calculated numerically, using an 8 level model (based on the levels depicted in Fig. \ref{fig:energylevel}), over a wide range of parameters. 
The rate equations dictating the populations for Fig. \ref{fig:greenSNR}, as well as for Fig. \ref{fig:IRSNR}, are the following: 

\begingroup\makeatletter\def\f@size{9}\check@mathfonts
\def\maketag@@@#1{\hbox{\m@th\large\normalfont#1}}%

\begin{align*}
&\dot{P}^-_{g,0} = -K^-_eP^-_{g,0}+K^-_fP^-_{e,0}+K^-_{sg,0}P_{s,g}+\frac{1}{2} (K{r_G}+K{r_{IR}})P^0_e \\ &\dot{P}^-_{g,1} = -K^-_eP^-_{g,1}+K^-_fP^-_{e,1}+K^-_{sg,1}P_{s,g}+\frac{1}{2} (K{r_G}+K{r_{{IR}}})P^0_e \\
&\dot{P}^-_{e,0} = -(K^-_f+K^-_{es,0}+K_{i_G}+K_{i_{IR}})P^-_{e,0}+K^-_eP_{g,0} \\
&\dot{P}^-_{e,1} = -(K^-_f+K^-_{es,1}+K_{i_G}+K_{i_{IR}})P^-_{e,1}+K^-_eP_{g,1} \\
&\dot{P}_{s,e} = -K^-_{ss}{P_{s,e}}+K^-_{es,0}P^-_{e,0}+K^-_{es,1}P^-_{e,1}+K^-_sP_{s,g} \\
&\dot{P}_{s,g} = -(K^-_{sg,0}+K^-_{sg,1})P_{s,g}-K^-_sP_{s,g}+K^-_{ss}{P_{s,e}} \\
&\dot{P}^0_g = -K^0_eP^0_g+K^0_fP^0_e+(K_{i_G}+K_{i_{IR}})(P^-_{e,0}+P^-_{e,1}) \\
&\dot{P}^0_e = -(K^0_f+K{r_G}+K{r_{{IR}}})P^0_e+K^0_eP^0_g.
\end{align*}\endgroup

In the above equations ${P}^-_{g,0}$ and ${P}^-_{g,1}$ represent the population in the $m_s = 0$ and $m_s = \pm 1$ triplet ground states of the negatively charged NV respectively, ${P}^-_{e,0}$ and ${P}^-_{e,1}$ represent the population in the $m_s = 0$ and $m_s = \pm 1$ triplet excited states of the negatively charged NV respectively, ${P}^0_g$ and ${P}^0_e$ represent the populations of the neutral charge NV ground and excited states respectively, and ${P}_{s,g}$ and ${P}_{s,e}$ represent the populations in the ground and excited singlet states of the negatively charged NV respectively. $K^-_e$ and $K^0_e$ represent the green laser induced excitation rates of NV$^-$ and NV$^0$ ground states to the excited states respectively, $K^-_s$ represents the IR laser induced excitation rate from the ground singlet state to the excited singlet state, $K^-_f$ and $K^0_f$ represent the fluorescence rate from the NV$^-$ and NV$^0$ excited states to their ground states respectively, $K_{ss}$ represents the decay rate of the excited singlet state to the ground singlet state, $K^-_{es,0}$ and $K^-_{es,1}$ represent the decay rates from the triplet excited states to the excited singlet state, respectively, $K^-_{sg,0}$ and $K^-_{sg,1}$ represent the decay rates from the ground singlet state to the NV$^-$ $m_s = 0$ and $m_s = \pm 1$ triplet ground states respectively, $K_{i_G}$ and $K_{i_{IR}}$ represent the green and IR excitation induced ionization rates respectively, and $K_{r_G}$ and $K_{r_{IR}}$ represent the green and IR excitation induced recombination rates respectively (see \cite{meirzada_negative_2017}). 

Figure \ref{fig:greenSNR} illustrates the achievable red fluorescence spin readout SNR, assuming 100\% collection and perfect detection without external noise sources (such as dark counts). Figures (a) and (b) depict the absolute SNR, described in Eq. \ref{SNR}, over a wide range of green excitation powers and readout durations. Figures (c) and (d) present the SNR for the same power and duration regimes normalized by the square root of the pulse duration in $\mu s$. 
The significant difference in SNR between bulk and surface NVs stems from differences in ionization cross section of the $^3E$ level. 

With the optimal parameters, the SNR rises slightly above 0.25 for bulk NVs and 0.22 for surface NVs. Thus, the red fluorescence spin readout demands a very high number of iterations before the spin state can be determined, even with perfect collection and detection, which limits the sensitivity and fidelity in NV based sensing and quantum information applications. 

We now detail the IR fluorescence based spin readout scheme. The pulsed sequence, depicted in Fig. \ref{fig:IRSNR}(a), starts with a short and strong green excitation, populating the singlet ground state ($^1E$). Next, a short delay (represented by $\tau$) is introduced in order to avoid undesired ionization from the excited triplet state, followed by a strong and long 980 nm pulse that excites the NV from the ground singlet state ($^1E$) to the singlet excited state ($^1A$) while collecting the emitted 1042 nm fluorescence. 
Due to the fact that the IR laser does not excite the triplet ground state, no mixing processes are expected, enabling a relatively long measurement. By carefully tuning the green laser pulse power and duration, the sequence can be repeated 3 times before significant mixing (via the singlet manifold or ionization/recombination processes) takes place, thus enhancing the signal. 

Despite the poor radiative coupling between the $^1A$ and $^1E$ levels, the fast decay rate from the $^1A$ state \cite{Ulbricht_Excited_state_lifetime_2018} together with the relatively long shelving time in the $^1E$ state \cite{acosta_optical_2010}, enable a large number of cycles before the NV decays back to the $^3A$ ground state without risking photo-ionization, allowing for a large enough number of photons to be collected during a single measurement, for high enough excitation powers. 

Figure \ref{fig:IRSNR} depicts the IR fluorescence spin readout SNR as a function of IR laser power and pulse duration of bulk and surface NVs, with delay duration $\tau$ = 10 ns (optimized with respect to the excited state lifetime). The laser power and pulse duration are scaled logarithmically in order to cover all of the relevant parameter space. Perfect collection and detection efficiencies are assumed for comparison with the results shown in Fig. \ref{fig:greenSNR}. We neglect IR induced ionization from the singlet state, for which the cross section is currently unknown (but assumed to be small), and consider a radiative to non-radiative coupling ratio of $1/1000$ \cite{acosta_optical_2010}. Figures \ref{fig:IRSNR}(b) and \ref{fig:IRSNR}(c) present the calculated absolute SNR for bulk and surface NVs, showing an expected significant enhancement of the spin state readout SNR compared to the red fluorescence spin readout scheme for high enough IR power. In addition, in this scheme the SNR grows monotonically with readout duration due to the absence of spin mixing. 
Figures \ref{fig:IRSNR}(d) and \ref{fig:IRSNR}(e) present the calculated normalized SNR for bulk and surface NVs for the IR fluorescence method, showing that the normalized SNR can reach higher values than that of the red fluorescence spin readout SNR for bulk and surface NVs, for strong excitation powers. 

To further improve the spin readout SNR shown in Fig. \ref{fig:IRSNR}, while reducing the necessary IR excitation power, we need to overcome the weak fluorescence signal resulting from the non-radiative nature of the $^1A \rightarrow ^1E$ decay.
Thus, we propose using optical/plasmonic antennas, hyperbolic-metamaterials (HMM) \cite{wolf_purcell-enhanced_2015,livneh_highly_2011,harats_full_2014} or a photonic crystal cavity \cite{wan_two-dimensional_2018,mouradian_rectangular_2017} to strengthen the radiative coupling between the $^1A$ and $^1E$ states and thus increase the singlet fluorescence signal. 

Photonic crystal structures with small mode volumes ($V \approx \lambda/n$) and high quality factors (high frequency-to-bandwidth ratio in the resonator \cite{haroche_cavity_1989})
are now within reach \cite{wan_two-dimensional_2018,mouradian_rectangular_2017}, and together with the relatively narrow IR fluorescence spectral width are expected to provide high Purcell factors, especially for nanodiamonds and diamond films, but also potentially for bulk diamonds. 

The Purcell factor, an enhancement of the spontaneous emission rate from the excited state due to radiative coupling \cite{haroche_cavity_1989}, depends on the quality factor and mode volume in the following way:

\begin{equation} 
    F_p = \frac{3}{4\pi^2}\Bigl(\frac{\lambda}{n}\Bigr)^3\frac{Q}{V}
\label{eq:purcell}
\end{equation}

where $\lambda$ represents the wavelength, $Q$ represents and quality factor, $n$ represents the refractive index and V represents the mode volume. In terms of the rate equations, the radiative part of the decay rate is multiplied by the Purcell factor. The fact that only approximately 0.1\% of the decay results in photon emission, holds great potential for enhancing the signal level and thus the SNR. In addition, the high emission directionality induced by a photonic crystal structure may dramatically increase the collection efficiency, and thus the number of photons detected. 

\begin{figure}[tbh]
\subfigure[]
{\includegraphics[width = 0.75 \linewidth]{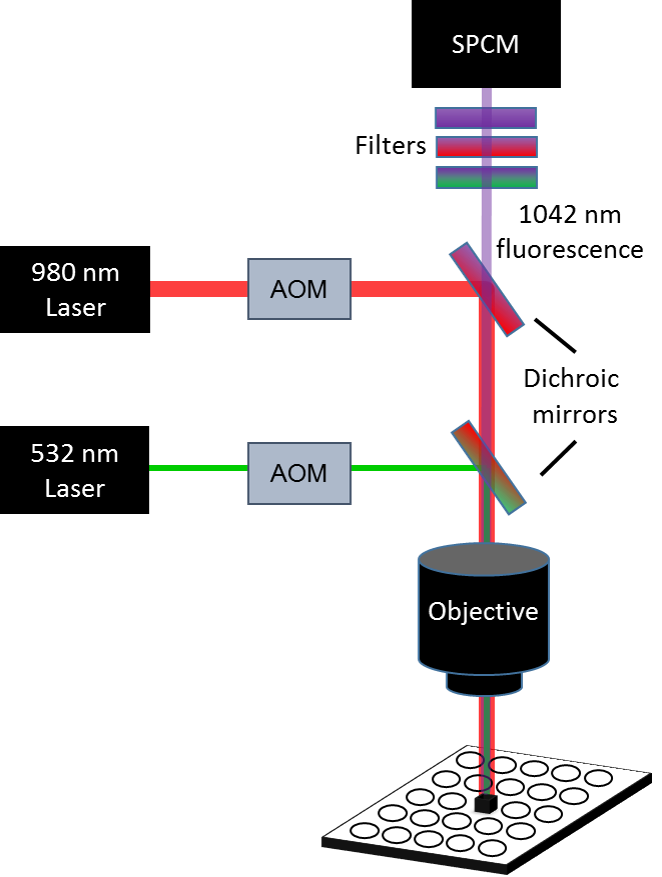}}
\subfigure[]
{\includegraphics[width = 0.46 \linewidth]{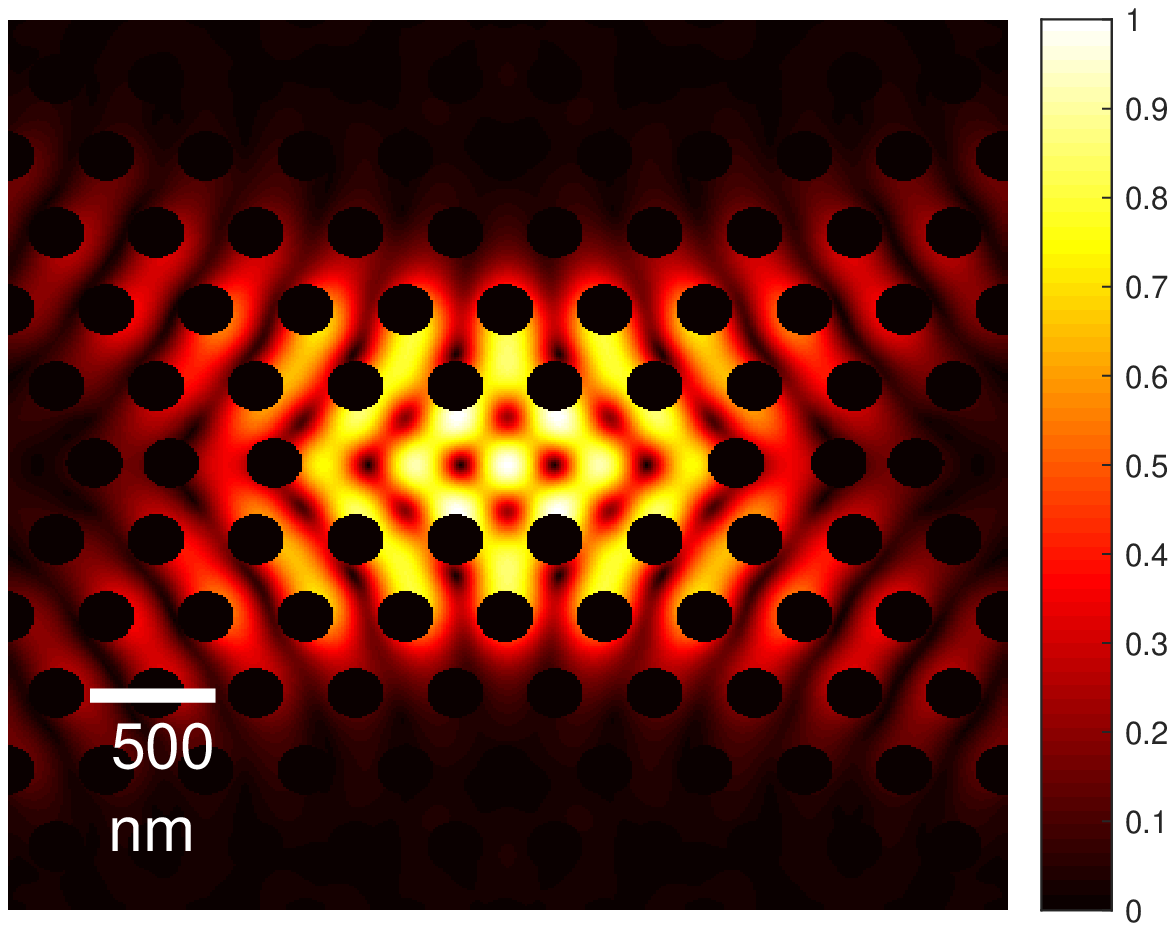}}
\subfigure[]
{\includegraphics[width = 0.46 \linewidth]{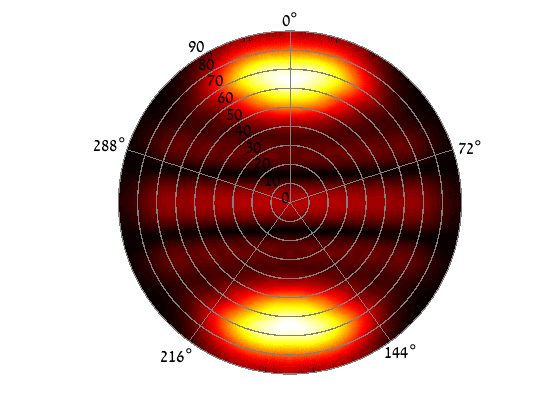}}
\caption{Schematic drawing of the suggested experimental setup and electric field energy density of the photonic crystal structure. (a) Schematic drawing of the suggested experimental setup. Green and IR lasers are focused onto the diamond sample in a photonic crystal cavity structure using a high NA objective lens, while AOMs control the pulse sequencing. Fluorescence is then collected through the same objective and directed to an SPCM after filtering the red fluorescence and the green and IR photons reflected from the diamond's surface. (b) Photonic crystal structure and electric field near-field energy density. 3 holes are shifted from each side to optimize Purcell factor. (c) Electric field far-field energy level density, showing highly directed emission from the cavity.}
\label{fig:PhotonicCrystal}
\end{figure}

\begin{figure}[tbh]
\subfigure[]
{\includegraphics[width = 0.99 \linewidth]{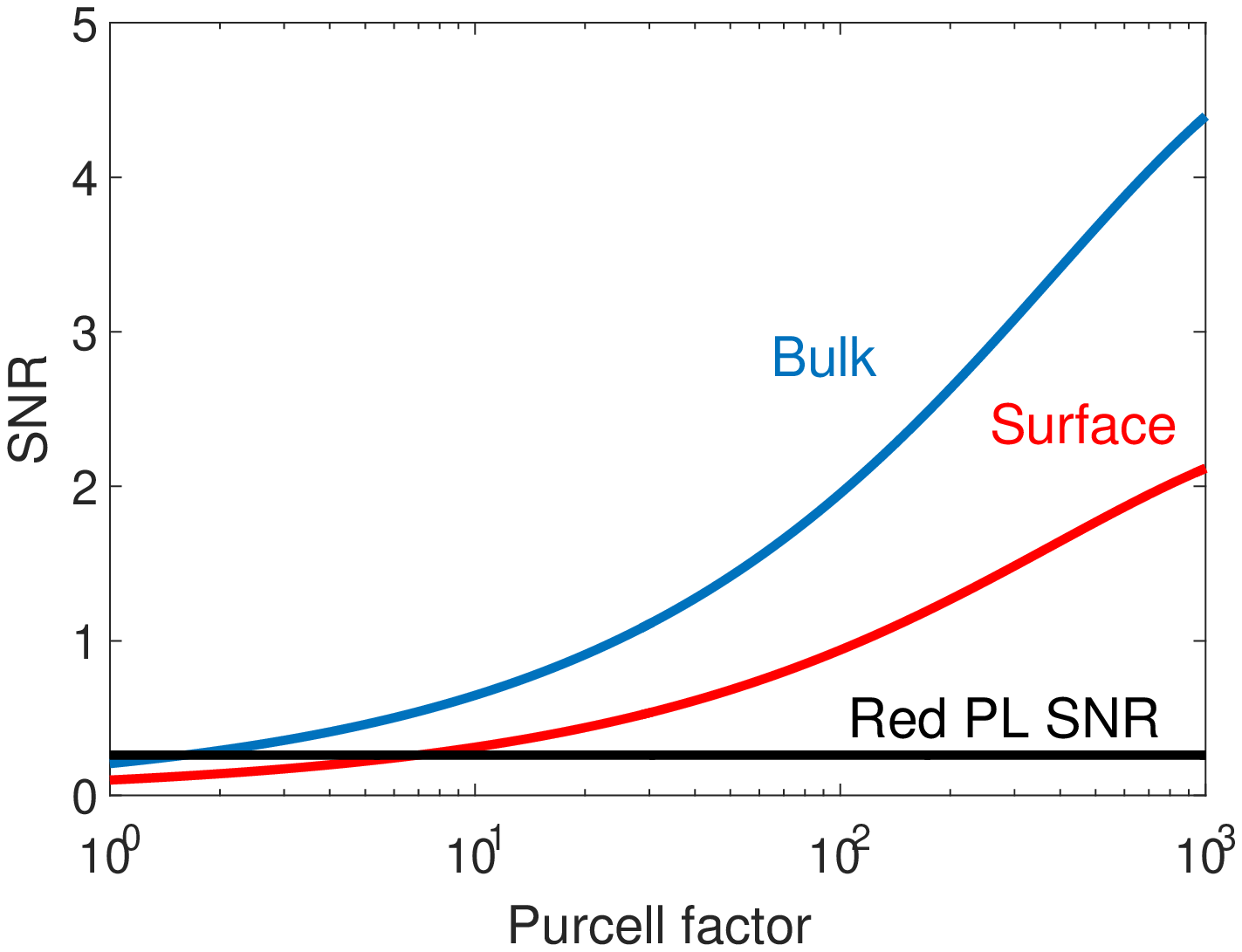}}
\subfigure[]
{\includegraphics[width = 0.46 \linewidth]{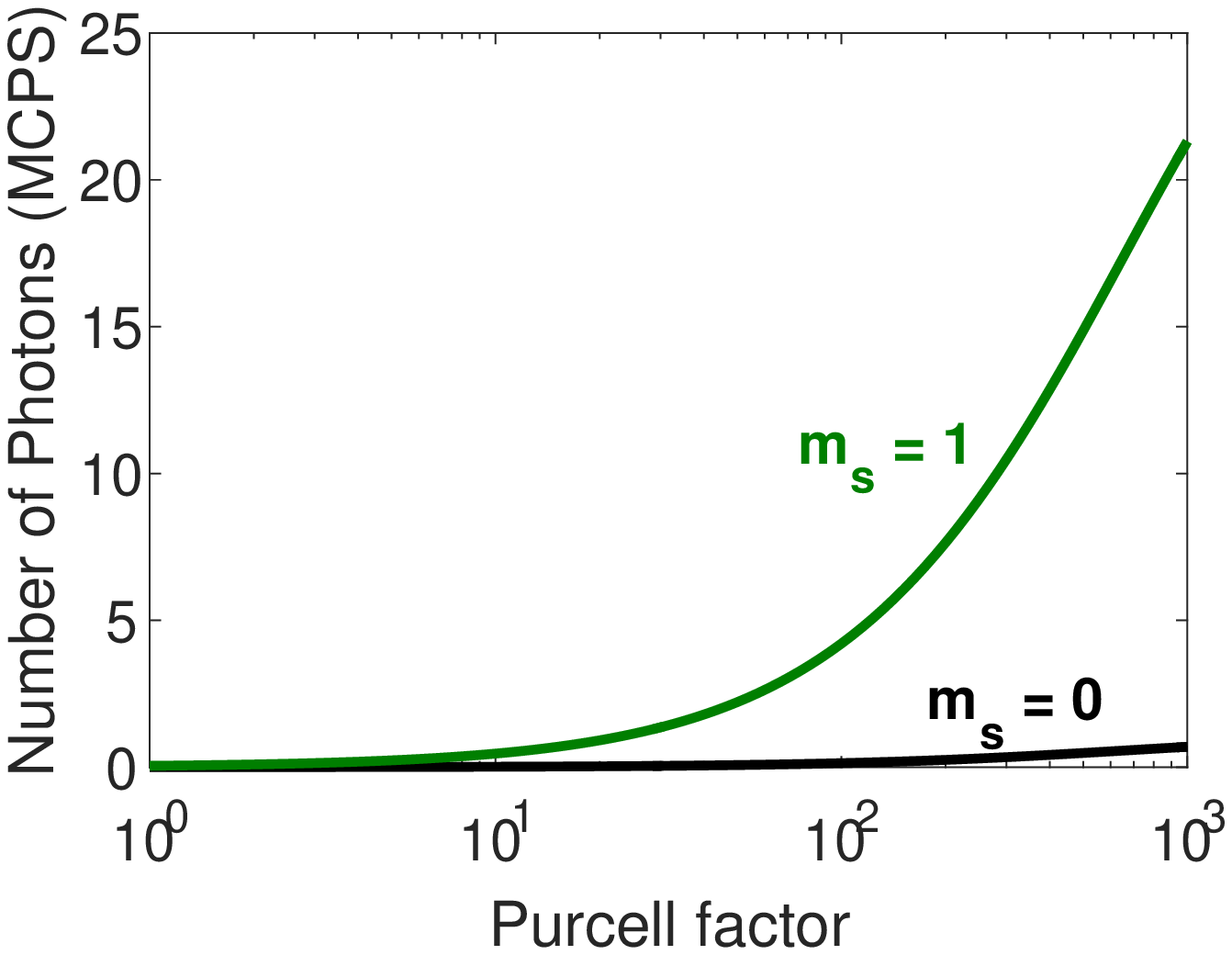}}
\subfigure[]
{\includegraphics[width = 0.46 \linewidth]{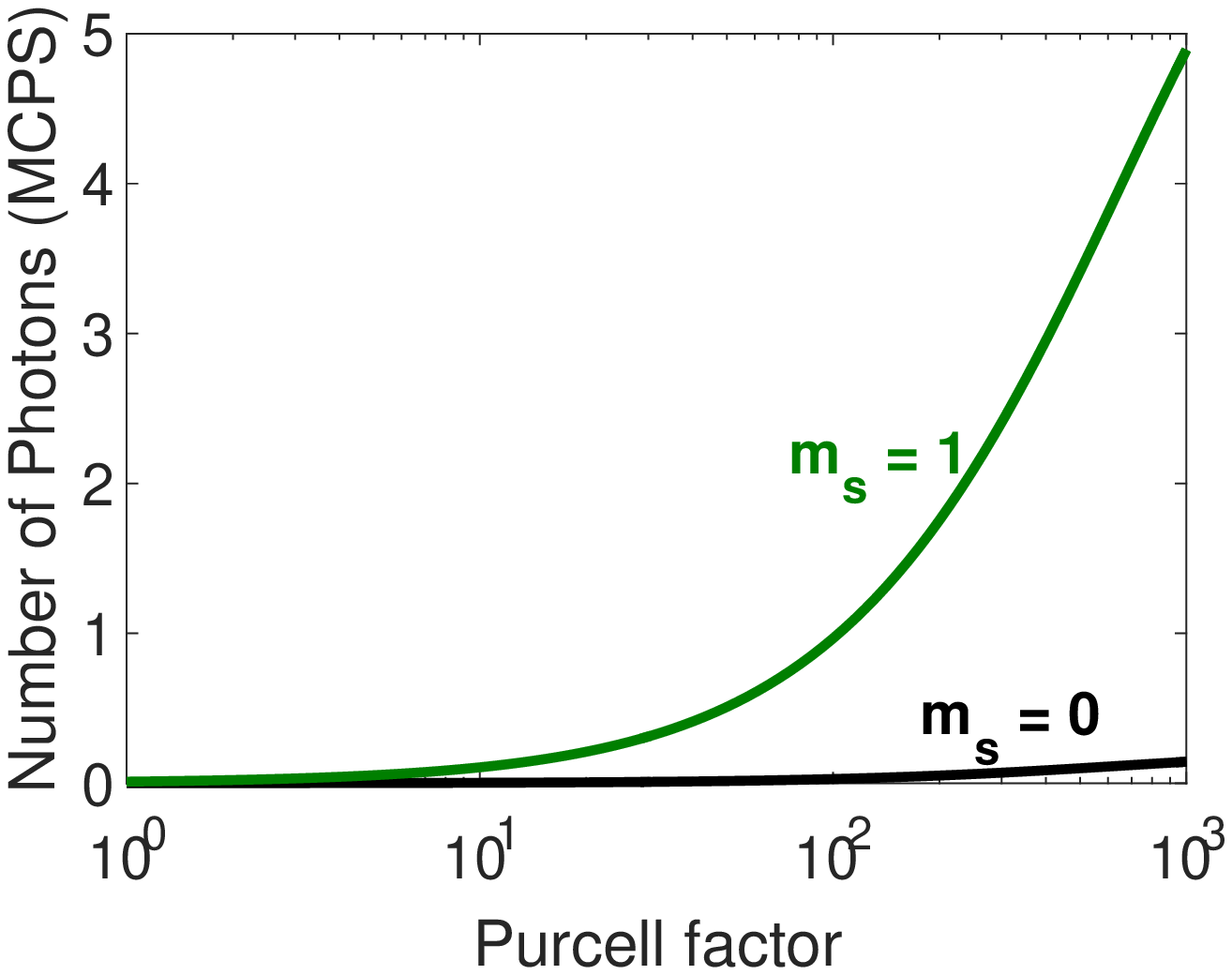}}
\caption{Expected spin state SNR and number of photons emitted as a function of Purcell factor under 1W excitation (inside the cavity), 1 $\mu s$ readout duration and an optimized delay duration $\tau$ = 10 ns. (a) spin state readout SNR for bulk (blue line) and surface (red line) NVs. The black line illustrates the highest SNR possible for bulk NV using red fluorescence. (b,c) number of photons emitted during the singlet excitation for $m_s = 0$ (black line) and $m_s = \pm 1$ (green line) for bulk (b) and surface (c) NVs.} 
\label{fig:coupling}
\end{figure}

Figure \ref{fig:coupling}(a) describes schematically the suggested experimental system. Green and detuned IR lasers excite the triplet ($^3A$) and singlet ($^1E$) ground states, respectively, while Acousto-Optic Modulators (AOMs) modulate them. Two dichroic mirrors with proper cutoff wavelengths (533 nm - 979 nm and 981 nm - 1041 nm for the green and IR lasers, respectively) direct the lasers onto the objective and enable fluorescence collection on a single-photon counter module (SPCM), after the unwanted red fluorescence and reflected green and IR lasers are filtered out. The objective focuses the light onto the diamond sample, here illustrated as a nanodiamond, to reach the high intensity IR excitation needed for driving the singlet transition efficiently. 
Figures \ref{fig:PhotonicCrystal}.b and \ref{fig:PhotonicCrystal}.c illustrate the electric field's near-field and far-field energy densities, as well as the photonic crystal cavity structure, optimized for nanodiamonds. The cavity structure is a 250 nm thick Silicone-Nitride hexagonal PHC L3 cavity with five neighbouring hole positions shifted, as described in \cite{minkov_automated_2014}. For this structure, the refractive index is 2, the lattice constant, $a$, is 450 nm and hole radius is 125 nm, and the positions of the holes were shifted by $0.315a$, $0.35a$, $0.118a$, $0.205a$ and $0.284a$. 
The far-field energy density enables approximately 45\% collection efficiency with numerical aperture of 0.95, while the near-field simulations predict a quality factor of about 2650 for this structure. Considering the small mode volume of this structure, $0.27(\frac{\lambda}{n})^3$, the resulting Purcell factor according to Eq. \ref{eq:purcell}, which is manifested by $K^-_S$ in Fig. \ref{fig:energylevel}, can reach up to 2343, and thus significantly enhance the emission and the number of photons collected. Similar calculations for diamond membranes and bulk diamonds predict quality factors of up to 13,300 and 790 with mode volumes of $0.38 (\frac{\lambda}{n})^3$ and $0.8(\frac{\lambda}{n})^3$, respectively, resulting in Purcell factors of up to 8355 for diamond membranes and 235 for bulk diamonds (see supplemental material). 

Figure \ref{fig:coupling} illustrates the expected spin readout SNR under 1W of IR excitation (inside the cavity) and a short readout duration (1 $\mu s$), as a function of Purcell factor for both surface (red line) and bulk (blue line) NVs. For this calculation, the Purcell factor was manifested by the radiative part of the rate $K^-_s$ in Fig. \ref{fig:energylevel}. Based on the figure, the new scheme provides a 5 fold enhancement of the spin readout SNR for a feasible Purcell factor of 40, which was already achieved for Silicon-Vacancy centers \cite{zhang_strongly_2018}, and more than an order of magnitude enhancement for $F_p = 300$ and $F_p = 1000$ (which are significantly lower than the Purcell factors calculated for nano-diamonds and diamond membranes) for bulk and surface NVs, respectively, thus exceeding the single-shot readout threshold. The SNR can reach even higher values for readout duration $>$ 1 $\mu s$ and higher excitation powers, as shown in the supplemental material. Thus, the magnetic field sensitivity, which obeys the following relation \cite{taylor_high-sensitivity_2008, Linh_thesis}: 

\begin{equation} 
\eta \propto \delta B \sqrt{T} \propto \frac{1}{SNR}
\label{eq:sensitivity}
\end{equation}

 could be reduced by more than an order of magnitude as well. 

Figures \ref{fig:coupling}(b) and \ref{fig:coupling}(c) present a calculation of the number of photons emitted from the $m_s = 0$ and $m_s = \pm1$ spin states as a function of Purcell factor for the same excitation power during the 1 $\mu s$ readout duration, showing that a higher number of photons is expected to be emitted during the readout sequence, while the contrast between the two spin states is sustained. 
\
\

In this work we presented a new spin state readout scheme, based on the IR fluorescence emitted from the singlet manifold following IR excitation from the singlet ground state. Using numerical calculations, we showed that this scheme results in more than two orders of magnitude enhancement of the spin readout SNR compared to the commonly used red fluorescence spin readout scheme. 
The NV center's singlet states were hardly addressed in NV center research \cite{acosta_optical_2010, PhysRevB.88.165202}, and so far few references described applications based on singlet excitation (\cite{dumeige_magnetometry_2013, shields_efficient_2015}). Our readout method complements the absorption based magnetometry presented in \cite{dumeige_magnetometry_2013}, as it generalizes it using the $^3E \rightarrow ^3A$ transitions for low concentrations of defects, enabling this transition to be used in a wider range of applications based on single or few NVs as well as for high density samples. 
Compared to other spin readout methods presented in recent years - spin-to-charge readout \cite{shields_efficient_2015,hopper_near-infrared-assisted_2016}, nuclear spin coupling \cite{steiner_universal_2010}, and resonant excitation(\cite{robledo_high-fidelity_2011} - our scheme could provide advantages in terms of measurement duration and the conditions required (1 $\mu$s readout in room temperature vs. 100 ms readout or ultra-cold systems). The significant SNR enhancement is expected to have a dramatic effect on nearly every NV based application currently pursued: the fact that only a few repetitions are needed (instead of the usual tens of thousands) will result in significantly improved sensitivities in magnetometry and strain sensing, as they are measured with respect to experiment duration. In addition, the enhanced SNR may remove a major obstacle in using NVs for quantum information processing, due to the importance of readout fidelity in this field. We are currently realizing this scheme experimentally, aiming to demonstrate enhanced spin readout SNR and improved magnetic field sensitivity. 


\bibliography{refs}

\begin{thebibliography}{28}%
\makeatletter
\providecommand \@ifxundefined [1]{%
 \@ifx{#1\undefined}
}%
\providecommand \@ifnum [1]{%
 \ifnum #1\expandafter \@firstoftwo
 \else \expandafter \@secondoftwo
 \fi
}%
\providecommand \@ifx [1]{%
 \ifx #1\expandafter \@firstoftwo
 \else \expandafter \@secondoftwo
 \fi
}%
\providecommand \natexlab [1]{#1}%
\providecommand \enquote  [1]{``#1''}%
\providecommand \bibnamefont  [1]{#1}%
\providecommand \bibfnamefont [1]{#1}%
\providecommand \citenamefont [1]{#1}%
\providecommand \href@noop [0]{\@secondoftwo}%
\providecommand \href [0]{\begingroup \@sanitize@url \@href}%
\providecommand \@href[1]{\@@startlink{#1}\@@href}%
\providecommand \@@href[1]{\endgroup#1\@@endlink}%
\providecommand \@sanitize@url [0]{\catcode `\\12\catcode `\$12\catcode
  `\&12\catcode `\#12\catcode `\^12\catcode `\_12\catcode `\%12\relax}%
\providecommand \@@startlink[1]{}%
\providecommand \@@endlink[0]{}%
\providecommand \url  [0]{\begingroup\@sanitize@url \@url }%
\providecommand \@url [1]{\endgroup\@href {#1}{\urlprefix }}%
\providecommand \urlprefix  [0]{URL }%
\providecommand \Eprint [0]{\href }%
\providecommand \doibase [0]{http://dx.doi.org/}%
\providecommand \selectlanguage [0]{\@gobble}%
\providecommand \bibinfo  [0]{\@secondoftwo}%
\providecommand \bibfield  [0]{\@secondoftwo}%
\providecommand \translation [1]{[#1]}%
\providecommand \BibitemOpen [0]{}%
\providecommand \bibitemStop [0]{}%
\providecommand \bibitemNoStop [0]{.\EOS\space}%
\providecommand \EOS [0]{\spacefactor3000\relax}%
\providecommand \BibitemShut  [1]{\csname bibitem#1\endcsname}%
\let\auto@bib@innerbib\@empty
\bibitem [{\citenamefont {Reed}\ \emph {et~al.}(2010)\citenamefont {Reed},
  \citenamefont {DiCarlo}, \citenamefont {Johnson}, \citenamefont {Sun},
  \citenamefont {Schuster}, \citenamefont {Frunzio},\ and\ \citenamefont
  {Schoelkopf}}]{reed_high-fidelity_2010}%
  \BibitemOpen
  \bibfield  {author} {\bibinfo {author} {\bibfnamefont {M.~D.}\ \bibnamefont
  {Reed}}, \bibinfo {author} {\bibfnamefont {L.}~\bibnamefont {DiCarlo}},
  \bibinfo {author} {\bibfnamefont {B.~R.}\ \bibnamefont {Johnson}}, \bibinfo
  {author} {\bibfnamefont {L.}~\bibnamefont {Sun}}, \bibinfo {author}
  {\bibfnamefont {D.~I.}\ \bibnamefont {Schuster}}, \bibinfo {author}
  {\bibfnamefont {L.}~\bibnamefont {Frunzio}}, \ and\ \bibinfo {author}
  {\bibfnamefont {R.~J.}\ \bibnamefont {Schoelkopf}},\ }\href {\doibase
  10.1103/PhysRevLett.105.173601} {\bibfield  {journal} {\bibinfo  {journal}
  {Physical Review Letters}\ }\textbf {\bibinfo {volume} {105}} (\bibinfo
  {year} {2010}),\ 10.1103/PhysRevLett.105.173601}\BibitemShut {NoStop}%
\bibitem [{\citenamefont {Myerson}\ \emph {et~al.}(2008)\citenamefont
  {Myerson}, \citenamefont {Szwer}, \citenamefont {Webster}, \citenamefont
  {Allcock}, \citenamefont {Curtis}, \citenamefont {Imreh}, \citenamefont
  {Sherman}, \citenamefont {Stacey}, \citenamefont {Steane},\ and\
  \citenamefont {Lucas}}]{myerson_high-fidelity_2008}%
  \BibitemOpen
  \bibfield  {author} {\bibinfo {author} {\bibfnamefont {A.~H.}\ \bibnamefont
  {Myerson}}, \bibinfo {author} {\bibfnamefont {D.~J.}\ \bibnamefont {Szwer}},
  \bibinfo {author} {\bibfnamefont {S.~C.}\ \bibnamefont {Webster}}, \bibinfo
  {author} {\bibfnamefont {D.~T.~C.}\ \bibnamefont {Allcock}}, \bibinfo
  {author} {\bibfnamefont {M.~J.}\ \bibnamefont {Curtis}}, \bibinfo {author}
  {\bibfnamefont {G.}~\bibnamefont {Imreh}}, \bibinfo {author} {\bibfnamefont
  {J.~A.}\ \bibnamefont {Sherman}}, \bibinfo {author} {\bibfnamefont {D.~N.}\
  \bibnamefont {Stacey}}, \bibinfo {author} {\bibfnamefont {A.~M.}\
  \bibnamefont {Steane}}, \ and\ \bibinfo {author} {\bibfnamefont {D.~M.}\
  \bibnamefont {Lucas}},\ }\href {\doibase 10.1103/PhysRevLett.100.200502}
  {\bibfield  {journal} {\bibinfo  {journal} {Physical Review Letters}\
  }\textbf {\bibinfo {volume} {100}} (\bibinfo {year} {2008}),\
  10.1103/PhysRevLett.100.200502}\BibitemShut {NoStop}%
\bibitem [{\citenamefont {Steiner}\ \emph {et~al.}(2010)\citenamefont
  {Steiner}, \citenamefont {Neumann}, \citenamefont {Beck}, \citenamefont
  {Jelezko},\ and\ \citenamefont {Wrachtrup}}]{steiner_universal_2010}%
  \BibitemOpen
  \bibfield  {author} {\bibinfo {author} {\bibfnamefont {M.}~\bibnamefont
  {Steiner}}, \bibinfo {author} {\bibfnamefont {P.}~\bibnamefont {Neumann}},
  \bibinfo {author} {\bibfnamefont {J.}~\bibnamefont {Beck}}, \bibinfo {author}
  {\bibfnamefont {F.}~\bibnamefont {Jelezko}}, \ and\ \bibinfo {author}
  {\bibfnamefont {J.}~\bibnamefont {Wrachtrup}},\ }\href {\doibase
  10.1103/PhysRevB.81.035205} {\bibfield  {journal} {\bibinfo  {journal}
  {Physical Review B}\ }\textbf {\bibinfo {volume} {81}} (\bibinfo {year}
  {2010}),\ 10.1103/PhysRevB.81.035205}\BibitemShut {NoStop}%
\bibitem [{\citenamefont {Morello}\ \emph {et~al.}(2010)\citenamefont
  {Morello}, \citenamefont {Pla}, \citenamefont {Zwanenburg}, \citenamefont
  {Chan}, \citenamefont {Tan}, \citenamefont {Huebl}, \citenamefont
  {Möttönen}, \citenamefont {Nugroho}, \citenamefont {Yang}, \citenamefont
  {van Donkelaar}, \citenamefont {Alves}, \citenamefont {Jamieson},
  \citenamefont {Escott}, \citenamefont {Hollenberg}, \citenamefont {Clark},\
  and\ \citenamefont {Dzurak}}]{morello_single-shot_2010}%
  \BibitemOpen
  \bibfield  {author} {\bibinfo {author} {\bibfnamefont {A.}~\bibnamefont
  {Morello}}, \bibinfo {author} {\bibfnamefont {J.~J.}\ \bibnamefont {Pla}},
  \bibinfo {author} {\bibfnamefont {F.~A.}\ \bibnamefont {Zwanenburg}},
  \bibinfo {author} {\bibfnamefont {K.~W.}\ \bibnamefont {Chan}}, \bibinfo
  {author} {\bibfnamefont {K.~Y.}\ \bibnamefont {Tan}}, \bibinfo {author}
  {\bibfnamefont {H.}~\bibnamefont {Huebl}}, \bibinfo {author} {\bibfnamefont
  {M.}~\bibnamefont {Möttönen}}, \bibinfo {author} {\bibfnamefont {C.~D.}\
  \bibnamefont {Nugroho}}, \bibinfo {author} {\bibfnamefont {C.}~\bibnamefont
  {Yang}}, \bibinfo {author} {\bibfnamefont {J.~A.}\ \bibnamefont {van
  Donkelaar}}, \bibinfo {author} {\bibfnamefont {A.~D.~C.}\ \bibnamefont
  {Alves}}, \bibinfo {author} {\bibfnamefont {D.~N.}\ \bibnamefont {Jamieson}},
  \bibinfo {author} {\bibfnamefont {C.~C.}\ \bibnamefont {Escott}}, \bibinfo
  {author} {\bibfnamefont {L.~C.~L.}\ \bibnamefont {Hollenberg}}, \bibinfo
  {author} {\bibfnamefont {R.~G.}\ \bibnamefont {Clark}}, \ and\ \bibinfo
  {author} {\bibfnamefont {A.~S.}\ \bibnamefont {Dzurak}},\ }\href {\doibase
  10.1038/nature09392} {\bibfield  {journal} {\bibinfo  {journal} {Nature}\
  }\textbf {\bibinfo {volume} {467}},\ \bibinfo {pages} {687} (\bibinfo {year}
  {2010})}\BibitemShut {NoStop}%
\bibitem [{\citenamefont {Fuchs}\ \emph {et~al.}(2011)\citenamefont {Fuchs},
  \citenamefont {Burkard}, \citenamefont {Klimov},\ and\ \citenamefont
  {Awschalom}}]{fuchs_quantum_2011}%
  \BibitemOpen
  \bibfield  {author} {\bibinfo {author} {\bibfnamefont {G.~D.}\ \bibnamefont
  {Fuchs}}, \bibinfo {author} {\bibfnamefont {G.}~\bibnamefont {Burkard}},
  \bibinfo {author} {\bibfnamefont {P.~V.}\ \bibnamefont {Klimov}}, \ and\
  \bibinfo {author} {\bibfnamefont {D.~D.}\ \bibnamefont {Awschalom}},\ }\href
  {\doibase 10.1038/nphys2026} {\bibfield  {journal} {\bibinfo  {journal}
  {Nature Physics}\ }\textbf {\bibinfo {volume} {7}},\ \bibinfo {pages} {789}
  (\bibinfo {year} {2011})}\BibitemShut {NoStop}%
\bibitem [{\citenamefont {Clevenson}\ \emph {et~al.}(2015)\citenamefont
  {Clevenson}, \citenamefont {Trusheim}, \citenamefont {Teale}, \citenamefont
  {Schröder}, \citenamefont {Braje},\ and\ \citenamefont
  {Englund}}]{clevenson_broadband_2015}%
  \BibitemOpen
  \bibfield  {author} {\bibinfo {author} {\bibfnamefont {H.}~\bibnamefont
  {Clevenson}}, \bibinfo {author} {\bibfnamefont {M.~E.}\ \bibnamefont
  {Trusheim}}, \bibinfo {author} {\bibfnamefont {C.}~\bibnamefont {Teale}},
  \bibinfo {author} {\bibfnamefont {T.}~\bibnamefont {Schröder}}, \bibinfo
  {author} {\bibfnamefont {D.}~\bibnamefont {Braje}}, \ and\ \bibinfo {author}
  {\bibfnamefont {D.}~\bibnamefont {Englund}},\ }\href {\doibase
  10.1038/nphys3291} {\bibfield  {journal} {\bibinfo  {journal} {Nature
  Physics}\ }\textbf {\bibinfo {volume} {11}},\ \bibinfo {pages} {393}
  (\bibinfo {year} {2015})}\BibitemShut {NoStop}%
\bibitem [{\citenamefont {Acosta}\ \emph
  {et~al.}(2010{\natexlab{a}})\citenamefont {Acosta}, \citenamefont {Bauch},
  \citenamefont {Jarmola}, \citenamefont {Zipp}, \citenamefont {Ledbetter},\
  and\ \citenamefont {Budker}}]{acosta_broadband_2010}%
  \BibitemOpen
  \bibfield  {author} {\bibinfo {author} {\bibfnamefont {V.~M.}\ \bibnamefont
  {Acosta}}, \bibinfo {author} {\bibfnamefont {E.}~\bibnamefont {Bauch}},
  \bibinfo {author} {\bibfnamefont {A.}~\bibnamefont {Jarmola}}, \bibinfo
  {author} {\bibfnamefont {L.~J.}\ \bibnamefont {Zipp}}, \bibinfo {author}
  {\bibfnamefont {M.~P.}\ \bibnamefont {Ledbetter}}, \ and\ \bibinfo {author}
  {\bibfnamefont {D.}~\bibnamefont {Budker}},\ }\href {\doibase
  10.1063/1.3507884} {\bibfield  {journal} {\bibinfo  {journal} {Applied
  Physics Letters}\ }\textbf {\bibinfo {volume} {97}},\ \bibinfo {pages}
  {174104} (\bibinfo {year} {2010}{\natexlab{a}})}\BibitemShut {NoStop}%
\bibitem [{\citenamefont {Dolde}\ \emph {et~al.}(2011)\citenamefont {Dolde},
  \citenamefont {Fedder}, \citenamefont {Doherty}, \citenamefont {Nöbauer},
  \citenamefont {Rempp}, \citenamefont {Balasubramanian}, \citenamefont {Wolf},
  \citenamefont {Reinhard}, \citenamefont {Hollenberg}, \citenamefont
  {Jelezko},\ and\ \citenamefont {Wrachtrup}}]{dolde_electric-field_2011}%
  \BibitemOpen
  \bibfield  {author} {\bibinfo {author} {\bibfnamefont {F.}~\bibnamefont
  {Dolde}}, \bibinfo {author} {\bibfnamefont {H.}~\bibnamefont {Fedder}},
  \bibinfo {author} {\bibfnamefont {M.~W.}\ \bibnamefont {Doherty}}, \bibinfo
  {author} {\bibfnamefont {T.}~\bibnamefont {Nöbauer}}, \bibinfo {author}
  {\bibfnamefont {F.}~\bibnamefont {Rempp}}, \bibinfo {author} {\bibfnamefont
  {G.}~\bibnamefont {Balasubramanian}}, \bibinfo {author} {\bibfnamefont
  {T.}~\bibnamefont {Wolf}}, \bibinfo {author} {\bibfnamefont {F.}~\bibnamefont
  {Reinhard}}, \bibinfo {author} {\bibfnamefont {L.~C.~L.}\ \bibnamefont
  {Hollenberg}}, \bibinfo {author} {\bibfnamefont {F.}~\bibnamefont {Jelezko}},
  \ and\ \bibinfo {author} {\bibfnamefont {J.}~\bibnamefont {Wrachtrup}},\
  }\href {\doibase 10.1038/nphys1969} {\bibfield  {journal} {\bibinfo
  {journal} {Nature Physics}\ }\textbf {\bibinfo {volume} {7}},\ \bibinfo
  {pages} {459} (\bibinfo {year} {2011})}\BibitemShut {NoStop}%
\bibitem [{\citenamefont {Taylor}\ \emph {et~al.}(2008)\citenamefont {Taylor},
  \citenamefont {Cappellaro}, \citenamefont {Childress}, \citenamefont {Jiang},
  \citenamefont {Budker}, \citenamefont {Hemmer}, \citenamefont {Yacoby},
  \citenamefont {Walsworth},\ and\ \citenamefont
  {Lukin}}]{taylor_high-sensitivity_2008}%
  \BibitemOpen
  \bibfield  {author} {\bibinfo {author} {\bibfnamefont {J.~M.}\ \bibnamefont
  {Taylor}}, \bibinfo {author} {\bibfnamefont {P.}~\bibnamefont {Cappellaro}},
  \bibinfo {author} {\bibfnamefont {L.}~\bibnamefont {Childress}}, \bibinfo
  {author} {\bibfnamefont {L.}~\bibnamefont {Jiang}}, \bibinfo {author}
  {\bibfnamefont {D.}~\bibnamefont {Budker}}, \bibinfo {author} {\bibfnamefont
  {P.~R.}\ \bibnamefont {Hemmer}}, \bibinfo {author} {\bibfnamefont
  {A.}~\bibnamefont {Yacoby}}, \bibinfo {author} {\bibfnamefont
  {R.}~\bibnamefont {Walsworth}}, \ and\ \bibinfo {author} {\bibfnamefont
  {M.~D.}\ \bibnamefont {Lukin}},\ }\href {\doibase 10.1038/nphys1075}
  {\bibfield  {journal} {\bibinfo  {journal} {Nature Physics}\ }\textbf
  {\bibinfo {volume} {4}},\ \bibinfo {pages} {810} (\bibinfo {year}
  {2008})}\BibitemShut {NoStop}%
\bibitem [{\citenamefont {Loretz}\ \emph {et~al.}(2014)\citenamefont {Loretz},
  \citenamefont {Pezzagna}, \citenamefont {Meijer},\ and\ \citenamefont
  {Degen}}]{loretz_nanoscale_2014}%
  \BibitemOpen
  \bibfield  {author} {\bibinfo {author} {\bibfnamefont {M.}~\bibnamefont
  {Loretz}}, \bibinfo {author} {\bibfnamefont {S.}~\bibnamefont {Pezzagna}},
  \bibinfo {author} {\bibfnamefont {J.}~\bibnamefont {Meijer}}, \ and\ \bibinfo
  {author} {\bibfnamefont {C.~L.}\ \bibnamefont {Degen}},\ }\href {\doibase
  10.1063/1.4862749} {\bibfield  {journal} {\bibinfo  {journal} {Applied
  Physics Letters}\ }\textbf {\bibinfo {volume} {104}},\ \bibinfo {pages}
  {033102} (\bibinfo {year} {2014})}\BibitemShut {NoStop}%
\bibitem [{\citenamefont {Trusheim}\ and\ \citenamefont
  {Englund}(2016)}]{trusheim_wide-field_2016}%
  \BibitemOpen
  \bibfield  {author} {\bibinfo {author} {\bibfnamefont {M.~E.}\ \bibnamefont
  {Trusheim}}\ and\ \bibinfo {author} {\bibfnamefont {D.}~\bibnamefont
  {Englund}},\ }\href {\doibase 10.1088/1367-2630/aa5040} {\bibfield  {journal}
  {\bibinfo  {journal} {New Journal of Physics}\ }\textbf {\bibinfo {volume}
  {18}},\ \bibinfo {pages} {123023} (\bibinfo {year} {2016})}\BibitemShut
  {NoStop}%
\bibitem [{\citenamefont {Wolf}\ \emph {et~al.}(2015)\citenamefont {Wolf},
  \citenamefont {Rosenberg}, \citenamefont {Rapaport},\ and\ \citenamefont
  {Bar-Gill}}]{wolf_purcell-enhanced_2015}%
  \BibitemOpen
  \bibfield  {author} {\bibinfo {author} {\bibfnamefont {S.~A.}\ \bibnamefont
  {Wolf}}, \bibinfo {author} {\bibfnamefont {I.}~\bibnamefont {Rosenberg}},
  \bibinfo {author} {\bibfnamefont {R.}~\bibnamefont {Rapaport}}, \ and\
  \bibinfo {author} {\bibfnamefont {N.}~\bibnamefont {Bar-Gill}},\ }\href
  {\doibase 10.1103/PhysRevB.92.235410} {\bibfield  {journal} {\bibinfo
  {journal} {Physical Review B}\ }\textbf {\bibinfo {volume} {92}} (\bibinfo
  {year} {2015}),\ 10.1103/PhysRevB.92.235410}\BibitemShut {NoStop}%
\bibitem [{\citenamefont {Shields}\ \emph {et~al.}(2015)\citenamefont
  {Shields}, \citenamefont {Unterreithmeier}, \citenamefont {de~Leon},
  \citenamefont {Park},\ and\ \citenamefont {Lukin}}]{shields_efficient_2015}%
  \BibitemOpen
  \bibfield  {author} {\bibinfo {author} {\bibfnamefont {B.}~\bibnamefont
  {Shields}}, \bibinfo {author} {\bibfnamefont {Q.}~\bibnamefont
  {Unterreithmeier}}, \bibinfo {author} {\bibfnamefont {N.}~\bibnamefont
  {de~Leon}}, \bibinfo {author} {\bibfnamefont {H.}~\bibnamefont {Park}}, \
  and\ \bibinfo {author} {\bibfnamefont {M.}~\bibnamefont {Lukin}},\ }\href
  {\doibase 10.1103/PhysRevLett.114.136402} {\bibfield  {journal} {\bibinfo
  {journal} {Physical Review Letters}\ }\textbf {\bibinfo {volume} {114}}
  (\bibinfo {year} {2015}),\ 10.1103/PhysRevLett.114.136402}\BibitemShut
  {NoStop}%
\bibitem [{\citenamefont {Robledo}\ \emph {et~al.}(2011)\citenamefont
  {Robledo}, \citenamefont {Childress}, \citenamefont {Bernien}, \citenamefont
  {Hensen}, \citenamefont {Alkemade},\ and\ \citenamefont
  {Hanson}}]{robledo_high-fidelity_2011}%
  \BibitemOpen
  \bibfield  {author} {\bibinfo {author} {\bibfnamefont {L.}~\bibnamefont
  {Robledo}}, \bibinfo {author} {\bibfnamefont {L.}~\bibnamefont {Childress}},
  \bibinfo {author} {\bibfnamefont {H.}~\bibnamefont {Bernien}}, \bibinfo
  {author} {\bibfnamefont {B.}~\bibnamefont {Hensen}}, \bibinfo {author}
  {\bibfnamefont {P.~F.~A.}\ \bibnamefont {Alkemade}}, \ and\ \bibinfo {author}
  {\bibfnamefont {R.}~\bibnamefont {Hanson}},\ }\href {\doibase
  10.1038/nature10401} {\bibfield  {journal} {\bibinfo  {journal} {Nature}\
  }\textbf {\bibinfo {volume} {477}},\ \bibinfo {pages} {574} (\bibinfo {year}
  {2011})}\BibitemShut {NoStop}%
\bibitem [{\citenamefont {Hopper}\ \emph {et~al.}(2016)\citenamefont {Hopper},
  \citenamefont {Grote}, \citenamefont {Exarhos},\ and\ \citenamefont
  {Bassett}}]{hopper_near-infrared-assisted_2016}%
  \BibitemOpen
  \bibfield  {author} {\bibinfo {author} {\bibfnamefont {D.~A.}\ \bibnamefont
  {Hopper}}, \bibinfo {author} {\bibfnamefont {R.~R.}\ \bibnamefont {Grote}},
  \bibinfo {author} {\bibfnamefont {A.~L.}\ \bibnamefont {Exarhos}}, \ and\
  \bibinfo {author} {\bibfnamefont {L.~C.}\ \bibnamefont {Bassett}},\ }\href
  {\doibase 10.1103/PhysRevB.94.241201} {\bibfield  {journal} {\bibinfo
  {journal} {Physical Review B}\ }\textbf {\bibinfo {volume} {94}} (\bibinfo
  {year} {2016}),\ 10.1103/PhysRevB.94.241201}\BibitemShut {NoStop}%
\bibitem [{\citenamefont {Meirzada}\ \emph {et~al.}(2018)\citenamefont
  {Meirzada}, \citenamefont {Hovav}, \citenamefont {Wolf},\ and\ \citenamefont
  {Bar-Gill}}]{meirzada_negative_2017}%
  \BibitemOpen
  \bibfield  {author} {\bibinfo {author} {\bibfnamefont {I.}~\bibnamefont
  {Meirzada}}, \bibinfo {author} {\bibfnamefont {Y.}~\bibnamefont {Hovav}},
  \bibinfo {author} {\bibfnamefont {S.~A.}\ \bibnamefont {Wolf}}, \ and\
  \bibinfo {author} {\bibfnamefont {N.}~\bibnamefont {Bar-Gill}},\ }\href
  {\doibase 10.1103/PhysRevB.98.245411} {\bibfield  {journal} {\bibinfo
  {journal} {Phys. Rev. B}\ }\textbf {\bibinfo {volume} {98}},\ \bibinfo
  {pages} {245411} (\bibinfo {year} {2018})}\BibitemShut {NoStop}%
\bibitem [{\citenamefont {Ulbricht}\ and\ \citenamefont
  {Loh}(2018)}]{Ulbricht_Excited_state_lifetime_2018}%
  \BibitemOpen
  \bibfield  {author} {\bibinfo {author} {\bibfnamefont {R.}~\bibnamefont
  {Ulbricht}}\ and\ \bibinfo {author} {\bibfnamefont {Z.-H.}\ \bibnamefont
  {Loh}},\ }\href {\doibase 10.1103/PhysRevB.98.094309} {\bibfield  {journal}
  {\bibinfo  {journal} {Phys. Rev. B}\ }\textbf {\bibinfo {volume} {98}},\
  \bibinfo {pages} {094309} (\bibinfo {year} {2018})}\BibitemShut {NoStop}%
\bibitem [{\citenamefont {Acosta}\ \emph
  {et~al.}(2010{\natexlab{b}})\citenamefont {Acosta}, \citenamefont {Jarmola},
  \citenamefont {Bauch},\ and\ \citenamefont {Budker}}]{acosta_optical_2010}%
  \BibitemOpen
  \bibfield  {author} {\bibinfo {author} {\bibfnamefont {V.~M.}\ \bibnamefont
  {Acosta}}, \bibinfo {author} {\bibfnamefont {A.}~\bibnamefont {Jarmola}},
  \bibinfo {author} {\bibfnamefont {E.}~\bibnamefont {Bauch}}, \ and\ \bibinfo
  {author} {\bibfnamefont {D.}~\bibnamefont {Budker}},\ }\href {\doibase
  10.1103/PhysRevB.82.201202} {\bibfield  {journal} {\bibinfo  {journal}
  {Physical Review B}\ }\textbf {\bibinfo {volume} {82}} (\bibinfo {year}
  {2010}{\natexlab{b}}),\ 10.1103/PhysRevB.82.201202}\BibitemShut {NoStop}%
\bibitem [{\citenamefont {Livneh}\ \emph {et~al.}(2011)\citenamefont {Livneh},
  \citenamefont {Strauss}, \citenamefont {Schwarz}, \citenamefont {Rosenberg},
  \citenamefont {Zimran}, \citenamefont {Yochelis}, \citenamefont {Chen},
  \citenamefont {Banin}, \citenamefont {Paltiel},\ and\ \citenamefont
  {Rapaport}}]{livneh_highly_2011}%
  \BibitemOpen
  \bibfield  {author} {\bibinfo {author} {\bibfnamefont {N.}~\bibnamefont
  {Livneh}}, \bibinfo {author} {\bibfnamefont {A.}~\bibnamefont {Strauss}},
  \bibinfo {author} {\bibfnamefont {I.}~\bibnamefont {Schwarz}}, \bibinfo
  {author} {\bibfnamefont {I.}~\bibnamefont {Rosenberg}}, \bibinfo {author}
  {\bibfnamefont {A.}~\bibnamefont {Zimran}}, \bibinfo {author} {\bibfnamefont
  {S.}~\bibnamefont {Yochelis}}, \bibinfo {author} {\bibfnamefont
  {G.}~\bibnamefont {Chen}}, \bibinfo {author} {\bibfnamefont {U.}~\bibnamefont
  {Banin}}, \bibinfo {author} {\bibfnamefont {Y.}~\bibnamefont {Paltiel}}, \
  and\ \bibinfo {author} {\bibfnamefont {R.}~\bibnamefont {Rapaport}},\ }\href
  {\doibase 10.1021/nl200052j} {\bibfield  {journal} {\bibinfo  {journal} {Nano
  Letters}\ }\textbf {\bibinfo {volume} {11}},\ \bibinfo {pages} {1630}
  (\bibinfo {year} {2011})}\BibitemShut {NoStop}%
\bibitem [{\citenamefont {Harats}\ \emph {et~al.}(2014)\citenamefont {Harats},
  \citenamefont {Livneh}, \citenamefont {Zaiats}, \citenamefont {Yochelis},
  \citenamefont {Paltiel}, \citenamefont {Lifshitz},\ and\ \citenamefont
  {Rapaport}}]{harats_full_2014}%
  \BibitemOpen
  \bibfield  {author} {\bibinfo {author} {\bibfnamefont {M.~G.}\ \bibnamefont
  {Harats}}, \bibinfo {author} {\bibfnamefont {N.}~\bibnamefont {Livneh}},
  \bibinfo {author} {\bibfnamefont {G.}~\bibnamefont {Zaiats}}, \bibinfo
  {author} {\bibfnamefont {S.}~\bibnamefont {Yochelis}}, \bibinfo {author}
  {\bibfnamefont {Y.}~\bibnamefont {Paltiel}}, \bibinfo {author} {\bibfnamefont
  {E.}~\bibnamefont {Lifshitz}}, \ and\ \bibinfo {author} {\bibfnamefont
  {R.}~\bibnamefont {Rapaport}},\ }\href {\doibase 10.1021/nl502652k}
  {\bibfield  {journal} {\bibinfo  {journal} {Nano Letters}\ }\textbf {\bibinfo
  {volume} {14}},\ \bibinfo {pages} {5766} (\bibinfo {year}
  {2014})}\BibitemShut {NoStop}%
\bibitem [{\citenamefont {Wan}\ \emph {et~al.}(2018)\citenamefont {Wan},
  \citenamefont {Mouradian},\ and\ \citenamefont
  {Englund}}]{wan_two-dimensional_2018}%
  \BibitemOpen
  \bibfield  {author} {\bibinfo {author} {\bibfnamefont {N.~H.}\ \bibnamefont
  {Wan}}, \bibinfo {author} {\bibfnamefont {S.}~\bibnamefont {Mouradian}}, \
  and\ \bibinfo {author} {\bibfnamefont {D.}~\bibnamefont {Englund}},\ }\href
  {\doibase 10.1063/1.5021349} {\bibfield  {journal} {\bibinfo  {journal}
  {Applied Physics Letters}\ }\textbf {\bibinfo {volume} {112}},\ \bibinfo
  {pages} {141102} (\bibinfo {year} {2018})}\BibitemShut {NoStop}%
\bibitem [{\citenamefont {Mouradian}\ \emph {et~al.}(2017)\citenamefont
  {Mouradian}, \citenamefont {Wan}, \citenamefont {Schröder},\ and\
  \citenamefont {Englund}}]{mouradian_rectangular_2017}%
  \BibitemOpen
  \bibfield  {author} {\bibinfo {author} {\bibfnamefont {S.}~\bibnamefont
  {Mouradian}}, \bibinfo {author} {\bibfnamefont {N.~H.}\ \bibnamefont {Wan}},
  \bibinfo {author} {\bibfnamefont {T.}~\bibnamefont {Schröder}}, \ and\
  \bibinfo {author} {\bibfnamefont {D.}~\bibnamefont {Englund}},\ }\href
  {\doibase 10.1063/1.4992118} {\bibfield  {journal} {\bibinfo  {journal}
  {Applied Physics Letters}\ }\textbf {\bibinfo {volume} {111}},\ \bibinfo
  {pages} {021103} (\bibinfo {year} {2017})}\BibitemShut {NoStop}%
\bibitem [{\citenamefont {Haroche}\ and\ \citenamefont
  {Kleppner}(1989)}]{haroche_cavity_1989}%
  \BibitemOpen
  \bibfield  {author} {\bibinfo {author} {\bibfnamefont {S.}~\bibnamefont
  {Haroche}}\ and\ \bibinfo {author} {\bibfnamefont {D.}~\bibnamefont
  {Kleppner}},\ }\href {\doibase 10.1063/1.881201} {\bibfield  {journal}
  {\bibinfo  {journal} {Physics Today}\ }\textbf {\bibinfo {volume} {42}},\
  \bibinfo {pages} {24} (\bibinfo {year} {1989})}\BibitemShut {NoStop}%
\bibitem [{\citenamefont {Minkov}\ and\ \citenamefont
  {Savona}(2014)}]{minkov_automated_2014}%
  \BibitemOpen
  \bibfield  {author} {\bibinfo {author} {\bibfnamefont {M.}~\bibnamefont
  {Minkov}}\ and\ \bibinfo {author} {\bibfnamefont {V.}~\bibnamefont
  {Savona}},\ }\href {https://doi.org/10.1038/srep05124} {\bibfield  {journal}
  {\bibinfo  {journal} {Scientific Reports}\ }\textbf {\bibinfo {volume} {4}},\
  \bibinfo {pages} {5124} (\bibinfo {year} {2014})}\BibitemShut {NoStop}%
\bibitem [{\citenamefont {Zhang}\ \emph {et~al.}(2018)\citenamefont {Zhang},
  \citenamefont {Sun}, \citenamefont {Burek}, \citenamefont {Dory},
  \citenamefont {Tzeng}, \citenamefont {Fischer}, \citenamefont {Kelaita},
  \citenamefont {Lagoudakis}, \citenamefont {Radulaski}, \citenamefont {Shen},
  \citenamefont {Melosh}, \citenamefont {Chu}, \citenamefont {Lončar},\ and\
  \citenamefont {Vučković}}]{zhang_strongly_2018}%
  \BibitemOpen
  \bibfield  {author} {\bibinfo {author} {\bibfnamefont {J.~L.}\ \bibnamefont
  {Zhang}}, \bibinfo {author} {\bibfnamefont {S.}~\bibnamefont {Sun}}, \bibinfo
  {author} {\bibfnamefont {M.~J.}\ \bibnamefont {Burek}}, \bibinfo {author}
  {\bibfnamefont {C.}~\bibnamefont {Dory}}, \bibinfo {author} {\bibfnamefont
  {Y.-K.}\ \bibnamefont {Tzeng}}, \bibinfo {author} {\bibfnamefont {K.~A.}\
  \bibnamefont {Fischer}}, \bibinfo {author} {\bibfnamefont {Y.}~\bibnamefont
  {Kelaita}}, \bibinfo {author} {\bibfnamefont {K.~G.}\ \bibnamefont
  {Lagoudakis}}, \bibinfo {author} {\bibfnamefont {M.}~\bibnamefont
  {Radulaski}}, \bibinfo {author} {\bibfnamefont {Z.-X.}\ \bibnamefont {Shen}},
  \bibinfo {author} {\bibfnamefont {N.~A.}\ \bibnamefont {Melosh}}, \bibinfo
  {author} {\bibfnamefont {S.}~\bibnamefont {Chu}}, \bibinfo {author}
  {\bibfnamefont {M.}~\bibnamefont {Lončar}}, \ and\ \bibinfo {author}
  {\bibfnamefont {J.}~\bibnamefont {Vučković}},\ }\href {\doibase
  10.1021/acs.nanolett.7b05075} {\bibfield  {journal} {\bibinfo  {journal}
  {Nano Letters}\ }\textbf {\bibinfo {volume} {18}},\ \bibinfo {pages} {1360}
  (\bibinfo {year} {2018})}\BibitemShut {NoStop}%
\bibitem [{\citenamefont {Pham}(2013)}]{Linh_thesis}%
  \BibitemOpen
  \bibfield  {author} {\bibinfo {author} {\bibfnamefont {L.}~\bibnamefont
  {Pham}},\ }\emph {\bibinfo {title} {Magnetic Field Sensing with
  Nitrogen-Vacancy Color Centers in Diamond.}},\ \href@noop {} {Ph.D. thesis},\
  \bibinfo  {school} {Harvard University} (\bibinfo {year} {2013})\BibitemShut
  {NoStop}%
\bibitem [{\citenamefont {Kehayias}\ \emph {et~al.}(2013)\citenamefont
  {Kehayias}, \citenamefont {Doherty}, \citenamefont {English}, \citenamefont
  {Fischer}, \citenamefont {Jarmola}, \citenamefont {Jensen}, \citenamefont
  {Leefer}, \citenamefont {Hemmer}, \citenamefont {Manson},\ and\ \citenamefont
  {Budker}}]{PhysRevB.88.165202}%
  \BibitemOpen
  \bibfield  {author} {\bibinfo {author} {\bibfnamefont {P.}~\bibnamefont
  {Kehayias}}, \bibinfo {author} {\bibfnamefont {M.~W.}\ \bibnamefont
  {Doherty}}, \bibinfo {author} {\bibfnamefont {D.}~\bibnamefont {English}},
  \bibinfo {author} {\bibfnamefont {R.}~\bibnamefont {Fischer}}, \bibinfo
  {author} {\bibfnamefont {A.}~\bibnamefont {Jarmola}}, \bibinfo {author}
  {\bibfnamefont {K.}~\bibnamefont {Jensen}}, \bibinfo {author} {\bibfnamefont
  {N.}~\bibnamefont {Leefer}}, \bibinfo {author} {\bibfnamefont
  {P.}~\bibnamefont {Hemmer}}, \bibinfo {author} {\bibfnamefont {N.~B.}\
  \bibnamefont {Manson}}, \ and\ \bibinfo {author} {\bibfnamefont
  {D.}~\bibnamefont {Budker}},\ }\href {\doibase 10.1103/PhysRevB.88.165202}
  {\bibfield  {journal} {\bibinfo  {journal} {Phys. Rev. B}\ }\textbf {\bibinfo
  {volume} {88}},\ \bibinfo {pages} {165202} (\bibinfo {year}
  {2013})}\BibitemShut {NoStop}%
\bibitem [{\citenamefont {Dumeige}\ \emph {et~al.}(2013)\citenamefont
  {Dumeige}, \citenamefont {Chipaux}, \citenamefont {Jacques}, \citenamefont
  {Treussart}, \citenamefont {Roch}, \citenamefont {Debuisschert},
  \citenamefont {Acosta}, \citenamefont {Jarmola}, \citenamefont {Jensen},
  \citenamefont {Kehayias},\ and\ \citenamefont
  {Budker}}]{dumeige_magnetometry_2013}%
  \BibitemOpen
  \bibfield  {author} {\bibinfo {author} {\bibfnamefont {Y.}~\bibnamefont
  {Dumeige}}, \bibinfo {author} {\bibfnamefont {M.}~\bibnamefont {Chipaux}},
  \bibinfo {author} {\bibfnamefont {V.}~\bibnamefont {Jacques}}, \bibinfo
  {author} {\bibfnamefont {F.}~\bibnamefont {Treussart}}, \bibinfo {author}
  {\bibfnamefont {J.-F.}\ \bibnamefont {Roch}}, \bibinfo {author}
  {\bibfnamefont {T.}~\bibnamefont {Debuisschert}}, \bibinfo {author}
  {\bibfnamefont {V.~M.}\ \bibnamefont {Acosta}}, \bibinfo {author}
  {\bibfnamefont {A.}~\bibnamefont {Jarmola}}, \bibinfo {author} {\bibfnamefont
  {K.}~\bibnamefont {Jensen}}, \bibinfo {author} {\bibfnamefont
  {P.}~\bibnamefont {Kehayias}}, \ and\ \bibinfo {author} {\bibfnamefont
  {D.}~\bibnamefont {Budker}},\ }\href {\doibase 10.1103/PhysRevB.87.155202}
  {\bibfield  {journal} {\bibinfo  {journal} {Physical Review B}\ }\textbf
  {\bibinfo {volume} {87}} (\bibinfo {year} {2013}),\
  10.1103/PhysRevB.87.155202}\BibitemShut {NoStop}%
\end{thebibliography}%


%

\end{document}